\documentclass[mathpazo]{cicp}

\usepackage{bm}

\begin{document}
%%%%% title : short title may not be used but TITLE is required.
% \title{TITLE}
% \title[short title]{TITLE}
\title{Accelerating $N$-body simulation of self-gravitating systems with limited first-order post-Newtonian approximation}

%%%%% author(s) :
% single author:
% \author[name in running head]{AUTHOR\corrauth}
% [name in running head] is NOT OPTIONAL, it is a MUST.
% Use \corrauth to indicate the corresponding author.
% Use \email to provide email address of author.
% \footnote and \thanks are not used in the heading section.
% Another acknowlegments/support of grants, state in Acknowledgments section
% \section*{Acknowledgments}
\author[T.~Tatekawa]{Takayuki Tatekawa\corrauth}
\address{Department of Social Design Engineering, National Institute of Technology, Kochi College, 200-1 Monobe-Otsu, Nankoku, Kochi 
783-8508, Japan\\
Research Institute of Science and Engineering, Waseda University, 3-4-1 Okubo, Shinjuku, Tokyo 169-8555, Japan \\
}
\email{{\tt tatekawa@akane.waseda.jp} (T.~Tatekawa)}

% multiple authors:
% Note the use of \affil and \affilnum to link names and addresses.
% The author for correspondence is marked by \corrauth.
% use \emails to provide email addresses of authors
% e.g. below example has 3 authors, first author is also the corresponding
%      author, author 1 and 3 having the same address.
% \author[Zhang Z R et.~al.]{Zhengru Zhang\affil{1}\comma\corrauth,
%       Author Chan\affil{2}, and Author Zhao\affil{1}}
% \address{\affilnum{1}\ School of Mathematical Sciences,
%          Beijing Normal University,
%          Beijing 100875, P.R. China. \\
%           \affilnum{2}\ Department of Mathematics,
%           Hong Kong Baptist University, Hong Kong SAR}
% \emails{{\tt zhang@email} (Z.~Zhang), {\tt chan@email} (A.~Chan),
%          {\tt zhao@email} (A.~Zhao)}
% \footnote and \thanks are not used in the heading section.
% Another acknowlegments/support of grants, state in Acknowledgments section
% \section*{Acknowledgments}

%%%%% Begin Abstract %%%%%%%%%%%
\begin{abstract}
In this study, an $N$-body simulation code was developed for self-gravitating systems with a limited
first-order post-Newtonian approximation. The code was applied to a special case
in which the system consists of one massive object and many low-mass objects.
Therefore, the behavior of stars around the massive black hole could be analyzed.
A graphics processing unit (GPU) was used to accelerate the code execution, and it could be accelerated by several tens of times compared
to a single-core CPU for $N \simeq 10^4$ objects.
\end{abstract}
%%%%% end %%%%%%%%%%%

%%%%% AMS/PACs/Keywords %%%%%%%%%%%
%\pac{}

%%%% maketitle %%%%%
\maketitle

%%%% Start %%%%%%
\section{Introduction}
\label{sec1}
The formation of massive black holes is one of the most important problems in astrophysics.
It was recently reported that a supermassive black hole (SMBH) exists at the center of the Milky Way 
~\cite{Schodel2002,Gillessen2009}. 
It has also been proposed that SMBHs, whose masses are estimated to be in the range of $10^6 - 10^{10} M_{\odot}$,
can exist in other galaxies,
according to the relation between the SMBH mass and the luminosity~\cite{Marconi2003},
or the SMBH mass and bulge mass~\cite{Haring2004}. In dwarf elliptical galaxies, the relation between massive black holes
and nuclear stars has been discussed~\cite{Graham2009}.

For globular clusters, the existence of massive black holes remains unclear. For example,
observations have indicated that the globular cluster NGC-224-G1 (or Mayall II) orbiting 
M31 can possess a massive black hole~\cite{Gebhardt2002,Gebhardt2005}.
In another case, the existence of a massive black hole in M15 (or NGC 7078) has also
been discussed~\cite{Gerssen2002,Baumgardt2003}. 
However, definitive conclusions have not yet been reached.

Although several scenarios for the formation of SMBHs have been discussed~\cite{Rees1984},
the exact scenario remains to be clearly identified. For example,
if the seeds of SMBHs were stellar-mass black holes, there would be an insufficient amount of time 
for them to grow into such a massive black hole.
The scenario where galaxies and SMBHs evolve together has also been discussed
~\cite{Kormendy2013}.

Gravity mainly affects the formation and evolution of astronomical objects.
Regarding dynamical evolution, $N$-body simulations have always been performed.
When we can see the effects of radiation or pressure of baryonic gas, we must consider the
hydrodynamical evolution. In contrast, we consider only the gravitational
interaction between objects where we can apply $N$-body simulations, in which
the interactions are described by Newtonian gravity. 

However, it is inadequate to describe the interaction of objects in neighboring regions of SMBHs by
Newtonian gravity alone. In Newtonian gravity, Bahcall and Wolf demonstrated that when a globular
cluster possesses a massive black hole, the density distribution peaks at the center
of the cluster~\cite{Bahcall1976}. Although their work is quite important,
when we discuss the behavior of stars near a massive black hole,
the effect of general relativity becomes important. Therefore,
we should consider the effect of general relativity in the neighboring regions of SMBHs.
For the $N$-body simulation, a post-Newtonian (PN) approach was proposed. 
The interaction of
massive objects is extended by $(v/c)^n$ terms. Then, the lowest-order term 
($(v/c)^2$) is
added to the Newtonian interaction. This equation of motion for $N$-body systems
is known as the EIH equation~\cite{Lorentz1937,Einstein1938}. The equations of motion
for $N$-body systems up to the second-order PN ($(v/c)^4$) have also been derived
~\cite{Ohta1973}.

In this study, a numerical simulation code for PN
$N$-body simulations was developed. Here, we note a special case, i.e., we suppose one SMBH
and many stars. The interaction between the SMBH and stars are estimated by the
PN approximation. Then, the interaction between stars is calculated
by Newtonian gravity. In this case, the procedure of the computation is reduced.
The interactions are computed on a graphics processing unit (GPU), which can
process a large number of operations in parallel. Because the computation
of complicated interactions is carried out on a GPU, the total computation time can be reduced.

The paper is organized as follows.
In Section \ref{sec:eq-of-motion}, the equation of motion and conserved quantities are described.
In a generic case, because of the emission of gravitational waves, the total energy of
the system decreases. In the first-order PN (1PN) approximation, because gravitational waves are not emitted, the total energy is conserved. Here, we note a
special case, i.e., the system consists of one massive object and many low-mass
stars. In Section \ref{sec:num}, the numerical simulation is described. Using a GPU,
the simulation can be accelerated. The elapsed time of the simulations are compared
between cases
of a central processing unit (CPU) only and a CPU+GPU. In Section 
\ref{sec:test_model}, the time evolution
for simple models is presented and the accuracy of the simulation is validated. Then,
the time evolution between Newtonian and PN cases is compared.
In Section \ref{sec:summary}, the conclusions of this study are presented.

\section{Equations of motion and conserved quantities}
\label{sec:eq-of-motion}
The equations of motion for $N$-body particles with the 1PN 
approximation were derived nearly 100 years ago~\cite{Lorentz1937,Einstein1938}.
These equations include the three-body interaction. Therefore,
when we consider $N$ particles, the order of computation
for the interaction becomes $O(N^3)$.
Because the computation of the interaction is computationally demanding,
the numerical simulation appears to be quite difficult.

Here, we assume that one object is considerably heavier than the other objects.
Under this assumption, the massive object only affects the relativistic correction.
In other words, we consider the interaction between the massive object
and other objects up to a 1PN approximation.
The interaction between the other objects are described only by Newtonian
gravity.
By this assumption, the order of computation
for the interaction decreases to $O(N^2)$~\cite{Will2014}.

Here, we label the massive object \#$1$. Then, the mass of particle \#$1$
is defined as $M$. The subscripts of the other objects are $i, j, k$.
The equation of motion for the massive object can be described as follows.
\begin{equation} \label{eqn:eq-object1}
\bm{a}_1 =  - \sum_j \frac{G m_j \bm{x}_{1j}}{r_{1j}^3}
 + \frac{1}{c^2} \left[ \bm{a}_1 \right ]_{BH} + \frac{1}{c^2}
  \left[ \bm{a}_1 \right ]_{Cross} + O \left ( \frac{G^2 m_j^3}{M c^2 r^3} \right )
  \,,
\end{equation}
where $m_j$ is the mass of object \#$j$ and
$\bm{x}_{1j} \equiv \bm{x}_j - \bm{x}_1, r_{1j} \equiv |\bm{x}_{1j} |$.
The sum over $j$ excludes
\#$1$. The first term of the right-hand side of Eq. (\ref{eqn:eq-object1})
represents Newtonian gravity. The PN terms in Eq. (\ref{eqn:eq-object1})
can be described as follows.
\begin{eqnarray}
\left [ \bm{a}_1 \right ]_{BH} &=& \sum_j \frac{G m_j \bm{x}_{1j}}{r_{1j}^3}
 \left ( 5 \frac{GM}{r_{1j}} - 2 v_j^2 + \frac{3}{2} \left ( \bm{v}_j \cdot 
  \bm{n}_{1j} \right ) \right ) \nonumber \\
 && + 3 \sum_j \frac{G m_j}{r_{1j}^3} (\bm{v}_j \cdot \bm{x}_{1j}) \bm{v}_j
  \,, \label{eqn:a1-BH} \\
\left[ \bm{a}_1 \right ]_{Cross} &=& 4 \sum_j \frac{G^2 m_j^2 \bm{x}_{1j}}{r_{1j}^4}
 \nonumber \\
 && + \sum_{j, k} \frac{G^2 m_j m_k \bm{x}_{1j}}{r_{1j}^3}
   \left ( \frac{4}{r_{1k}} + \frac{5}{4 r_{jk}} - \frac{r_{1k}^2}{4 r_{jk}^3}
    + \frac{r_{1j}^2}{4 r_{jk}^3} \right ) \nonumber \\
 && - \frac{7}{2} \sum_{j, k} \frac{G^2 m_j m_k \bm{x}_{jk}}{r_{jk}^3 r_{1j}}
  \nonumber \\
 && - \sum_{j, k} \frac{G m_j m_k}{M r_{1j}^3} 
   \left [ 4 (\bm{v}_j \cdot \bm{v}_k ) \bm{x}_{1j} - 3 ( \bm{v}_j \cdot \bm{x}_{1j}
    ) \bm{v}_k - 4 ( \bm{v}_k \cdot \bm{x}_{1j} ) \bm{v}_j \right ] \,.
    \label{eqn:a1-cross}
\end{eqnarray}
The sum over $k$ excludes \#$1$. $\bm{n}$ is the unit vector
$\bm{n}_{1j}=\bm{x}_{1j}/r_{1j}$.

The equation of motion for the low-mass stars can be described as follows.
\begin{equation} \label{eqn:eq-light_object}
\bm{a}_i =  -\frac{G M \bm{x}_{i1}}{r_{i1}^3}
 - \sum_j \frac{G m_j \bm{x}_{ij}}{r_{ij}}
 + \frac{1}{c^2} \left[ \bm{a}_i \right ]_{BH} + \frac{1}{c^2}
  \left [ \bm{a}_i \right ]_{Cross} + O \left ( \frac{G^2 m_j^2}{c^2 r^3} \right )
  \,,
\end{equation}
where
\begin{eqnarray}
\left[ \bm{a}_i \right ]_{BH} &=&
 \frac{GM \bm{x}_{i1}}{r_{i1}^3} \left ( 4 \frac{GM}{r_{i1}} - v_i^2
  \right ) + 4 \frac{GM}{r_{i1}^3}
   \left (\bm{v}_i \cdot \bm{x}_{i1} \right ) \bm{v}_a \,, 
   \label{eqn:ai-all} \\
\left [ \bm{a}_i \right ]_{Cross}
 &=& 5 \frac{G^2 m_i M \bm{x}_{i1}}{r_{i1}^4}
  - \frac{G m_i}{r_{i1}^3} \left [ 
   4 v_i^2 \bm{x}_{i1} - 7 (\bm{v}_i \cdot \bm{x}_{i1}) \bm{v}_i \right ]
   \nonumber \\
 && + \sum_j \frac{G^2 m_j M \bm{x}_{i1}}{r_{i1}^3}
  \left ( \frac{4}{r_{ij}} + \frac{5}{4 r_{j1}} + \frac{r_{i1}^2}{4 r_{j1}^3}
   - \frac{r_{ij}^2}{4 r_{j1}^3} \right ) \nonumber \\
 && + \sum_j \frac{G^2 m_j M \bm{x}_{ij}}{r_{ij}^3}
  \left ( \frac{4}{r_{i1}} + \frac{5}{4 r_{j1}} - \frac{r_{i1}^2}{4 r_{j1}^3}
   + \frac{r_{ij}^2}{4 r_{j1}^3} \right ) \nonumber \\
 && - \frac{7}{2} \sum_j \frac{G^2 m_j M \bm{x}_{j1}}{r_{j1}^3}
  \left ( \frac{1}{r_{ij}} - \frac{1}{r_{i1}} \right ) \nonumber \\
 && - \sum_j \frac{G m_j}{r_{i1}^3}
  \left [ 4 (\bm{v}_i \cdot \bm{v}_j) \bm{x}_{i1}
   - 3 (\bm{v}_j \cdot \bm{x}_{i1} ) \bm{v}_i
   - 4 (\bm{v}_i \cdot \bm{x}_{i1} ) \bm{v}_j \right ] \nonumber \\
 && + \sum_j \frac{G m_j \bm{x}_{ij}}{r_{ij}^3}
  \left [ v_i^2 - 2 | \bm{v}_{ij} |^2 + \frac{3}{2} ( \bm{v}_j \cdot
   \bm{n}_{ij} )^2 \right ] \nonumber \\
 && + \sum_j \frac{G m_j}{r_{ij}^3} \left [ \bm{x}_{ij} \cdot
  ( 4 \bm{v}_i - 3 \bm{v}_j ) \right ] \bm{v}_{ij} \,, \label{eqn:ai-cross}
\end{eqnarray}
where $\bm{v}_{ij} \equiv \bm{v}_j - \bm{v}_i$.

The total energy and $3$-momentum in the 1PN order are conserved. In the present study,
we evaluate the total energy during the time evolution.
\begin{eqnarray}
E &=& \frac{1}{2} M v_1^2 + \frac{1}{2} \sum_i m_i v_i^2 
 - \frac{1}{2} \sum_{i, j} \frac{G m_i m_j}{r_{ij}}
 - \sum_i \frac{G M m_i}{r_{1i}} \nonumber \\
&& + \frac{1}{c^2} \left \{ \frac{3}{8} \sum_i m_i v_i^4
 + \frac{3}{2} \sum_i \frac{GM m_i}{r_{1i}} v_i^2
 + \frac{1}{2} \sum_i \frac{G^2 M^2 m_i}{r_{1i}^2} \right . \nonumber \\
&& ~~~~ + \frac{1}{4} \sum_{i, j} \frac{G m_i m_j}{r_{ij}} \left [ 6 v_i^2
 - 7 \bm{v}_i \cdot \bm{v}_j - (\bm{n}_{ij} \cdot \bm{v}_i)
  (\bm{n}_{ij} \cdot \bm{v}_j) \right ]  \nonumber \\
&& ~~~~ + \frac{1}{2} \sum_i \frac{G M m_i}{r_{1i}} \left [ 3 v_1^2
 - 7 \bm{v}_1 \cdot \bm{v}_i
  - (\bm{n}_{1i} \cdot \bm{v}_1) (\bm{n}_{1i} \cdot \bm{v}_i) \right ] 
  \nonumber \\
&& ~~~~ \left . + \sum_{i,j} \frac{G^2 M m_i m_j}{r_{ij} r_{1i}}
 + \frac{1}{2} \sum_{i,j} \frac{G^2 M m_i m_j}{r_{1i} r_{1j}} \right \}
  + O \left ( \frac{G^2 m_i^3}{r^2} \right ) \,, 
   \label{eqn:energy} \\
\bm{P} &=& M \bm{v}_1 + \sum_i m_i \bm{v}_i 
 \left ( 1+ \frac{1}{2c^2} v_i^2 \right ) - \frac{1}{2c^2}
 \left \{ \sum_i \frac{GM m_i}{r_{1i}} \left [ \bm{v}_i + (\bm{v}_i \cdot \bm{n}_{1i} )
  \bm{n}_{1i} \right ] \right . \nonumber \\
&& \left . + \sum_i \frac{GM m_i}{r_{1i}} [ \bm{v}_1
 + (\bm{v}_1 \cdot \bm{n}_{1i} ) \bm{n}_{1i} ]
 + \sum_{i,j} \frac{G m_i m_j}{r_{ij}} [ \bm{v}_i
 + (\bm{v}_i \cdot \bm{n}_{ij} ) \bm{n}_{ij} ] \right \} 
 \nonumber \\
&& + O \left ( \frac {G m_i^3 v_i}{c^2 M r} \right ) \,.
\end{eqnarray}

\section{Numerical Simulation}
\label{sec:num}
\subsection{Implementation for GPU}
\label{subsec:GPU}
For the acceleration of the numerical simulation, a GPU is used
for heavy calculations. For self-gravitating systems, the GRAvity piPE (GRAPE) system
was developed~\cite{GRAPE}. 
The GRAPE processor calculates the acceleration and gravitational potential
for each particle from the position and mass of particles.
In the case of $N$ particles, the calculation of the acceleration and
potential becomes $O(N^2)$.
Unfortunately, the GRAPE system can only be applied for Newtonian gravity.
In contrast, 
because the GPU is programmable, we can perform a generic computation using the GPU.

Computing platforms for the development of the GPU computation have been produced, e.g., NVIDIA CUDA~\cite{CUDA} and OpenCL~\cite{OpenCL}.
Although these environments are required to
describe the data transfer between the main memory and memory beside GPU,
it is difficult to describe the optimized code. Instead of these environments,
a command is introduced for the computation on the GPU to the source code. 
Here, we have applied the domain-specific compiler ``Goose,'' which was developed
by K \& F Computing Research~\cite{goose}. The target loops to compute are specified
on the GPU by the command. The syntax of the commands on ``Goose'' is similar 
to OpenACC~\cite{OpenACC}. Although ``Goose'' can be applied up to double loops,
this compiler can implement the reduction command for
array variables (for example, $\left[ \bm{a}_i \right ]_{BH}$) implicitly. Therefore, the computation
of the acceleration can be
accelerated easily.
In terms of three-body interactions, the loops of the variables $j$ and $k$ expand
to a significant number of threads and are executed on the GPU simultaneously.

For acceleration, we should choose the terms where the computation is significantly heavier
than the input or output data transfer. In other words, when the procedure of
the computation and data transfer are $O(N^2)$ and $O(N)$, respectively,
it is easy to use the GPU to accelerate it.
In our case, we calculate the terms of the sum over $j, k$ in Eq.~(\ref{eqn:a1-cross}) and
the terms of the sum over $j$ in Eqs.~(\ref{eqn:eq-light_object}) and
 (\ref{eqn:ai-cross}) on the GPU. Then, to evaluate the accuracy, the total
energy is obtained by calculating the terms of the sum over $i, j$ in
Eq.~(\ref{eqn:energy}). 

\subsection{Evaluation of acceleration}
\label{subsec:evaluation}
For the evaluation of acceleration, the numerical code is executed on a generic PC with a
GPU. The specifications of the PC are summarized in Table 3.1.

\begin{table}[h]
Table 3.1~ 
Specifications of PC.\\
\begin{tabular}{lr} \hline \hline
Instrument & model/version/quantity \\ \hline
CPU & Intel Core i7-3770K \\
RAM & 32 GB \\
OS & CentOS 6.9 \\
kernel & Version 2.6.32-642.11.1 \\
gcc & Version 4.4.7 \\
GPU & NVIDIA Tesla K20c\\
NVIDIA CUDA & Version 4.2 \\ 
Goose & Version 1.3.3 \\ \hline
\end{tabular}
%\label{tab:PC}
\end{table}

\begin{figure}[htb]
\centerline{
\includegraphics[height=7cm]{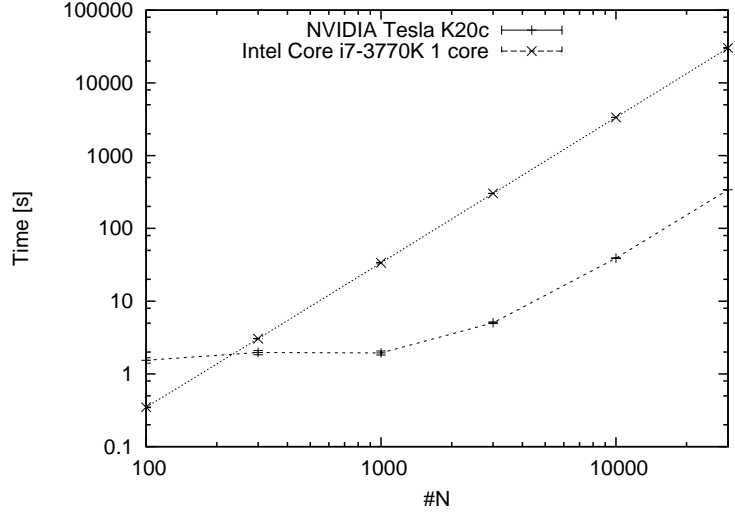}
}
\vspace{2cm}
\caption{
Dependency of the number of the particles
on the computation time. Because the dispersion is quite small,
we do not plot error bars.
In the case of the host CPU only, the computation time appears to be
proportional to $N^2$. In contrast, the computation time
is suppressed by the computation on the GPU by a large number of threads.
When $N$ increases to $10^4$,
the computation time increases because the threads of the GPU become
saturated.}
\label{fig:time-N}
\end{figure}

Here, the number of particles $N$ are varied in the range of
$[100, 30000]$ and the elapsed time is measured. 
Table 3.2 and
Figure~\ref{fig:time-N} show the dependency of the number of particles
for the total elapsed time. We calculate 32 time steps of time evolution 
with the fourth-order Runge--Kutta method~\cite{NR}, then calculate the total energy
at both the start and end of the computation.

\begin{table}[h]
Table 3.2~
Dependency of the number of particles
for the total elapsed time. Here, we take the average of 10 samples for each case.\\
\begin{tabular}{rrrrr} \hline \hline
\#$N$ & $t_{GPU}$ [s] & $\sigma_{t_{GPU}}$ [s] & $t_{CPU}$ [s] & $\sigma_{t_{CPU}}$ [s] \\
\hline
$100$ & $1.5392$ & $0.1299$ & $0.3479$ & $0.0043$ \\
$300$ & $1.9651$ & $0.1308$ & $3.0501$ & $0.0060$ \\
$1000$ & $1.941$ & $0.1222$ & $33.652$ & $0.0881$ \\
$3000$ & $5.0517$ & $0.1284$ & $302.31$ & $0.10062$ \\
$10000$ & $38.934$ & $0.0100$ & $3358.6$ & $2.1313$ \\
$30000$ & $339.956$ & $0.1391$ & $30188$ & $17.482$ \\ \hline
\end{tabular}
%\label{tab:time-N}
\end{table}

In the case of computation on the
host CPU only, the elapsed time is almost proportional to $O(N^2)$.
In contrast, in the case of computation with the GPU,
because multiple threads are used, the increase in computation
time is suppressed. However, when $N$ increases to $10^4$,
the loops of the three-body interactions expand
to $O(10^8)$ threads. Therefore, the threads of the GPU seem
saturated.

%%%% Figure %%%
%%%%%%%%%%%%%%

\section{Time evolution of test model}
\label{sec:test_model}
In a previous study, we analyzed the effect of a central massive object on
low-mass stars with Newtonian gravity~\cite{Tashiro2010,Tashiro2011}.
In the collisionless case, an explicit symplectic integrator conserves the 
total energy for a significant length of time in Newtonian self-gravitational systems
\cite{Suzuki1992,Yoshida1990,Yoshida1993}.
In contrast,
although we can describe the Hamiltonian for this model, we cannot apply the
explicit symplectic integrator, because the Hamiltonian cannot be divided into
coordinate terms and momentum terms in the 1PN equations.
Therefore, we apply an
alternative integrator. The terms of the 1PN appear to be
complicated; therefore, we avoid computing the time derivative of these terms.

In this study, we apply the fourth-order Runge--Kutta integration method.
Unlike the symplectic integrator, the Runge--Kutta method does not conserve the
total energy in the global timescale.

In the simulation, we set the constant $c=G=1$. Then, the total mass of low-mass stars
is set as
\begin{equation}
\sum_{i \ne 1} m_i = 1 \,.
\end{equation}
In this case, the Schwarzschild radius of the massive object becomes
$2G M/c^2 = 2 M$. We expect that
the effect of general relativity appears only around the Schwarzschild radius 
of the massive object.

The initial condition of the test model is given by a spherically symmetric distribution.
The low-mass stars are distributed in the region for $1 < r < 10$.
The spatial distribution is generated by a random number from
linear congruential generators~\cite{NR}.
The total number of low-mass stars is $N=10000$. Then, we assume that
all the low-mass stars have equal mass.
In other words, the mass of the low-mass stars
is given by $m=1/N=10^{-4}$. Then, the massive object $M=100 m$ is set at the origin.
The Schwarzschild radius of the massive object becomes $R_{Sch} = 0.02$.
The initial velocity of both the massive object and low-mass stars is set to zero.
In other words, we suppose a cold collapse of the system.

To avoid the dispersion of the interaction at the collision, we introduce
Plummer softening to the inverse of the distance between stars.
\begin{equation}
\frac{1}{r} \rightarrow \frac{1}{\sqrt{r^2 + \varepsilon^2}} \,.
\end{equation}
In this study, we set the softening parameter $\varepsilon= 10^{-3}$, which is
shorter than the Schwarzschild radius of the massive object.

\begin{figure}[htb]
\centerline{
\includegraphics[height=12cm]{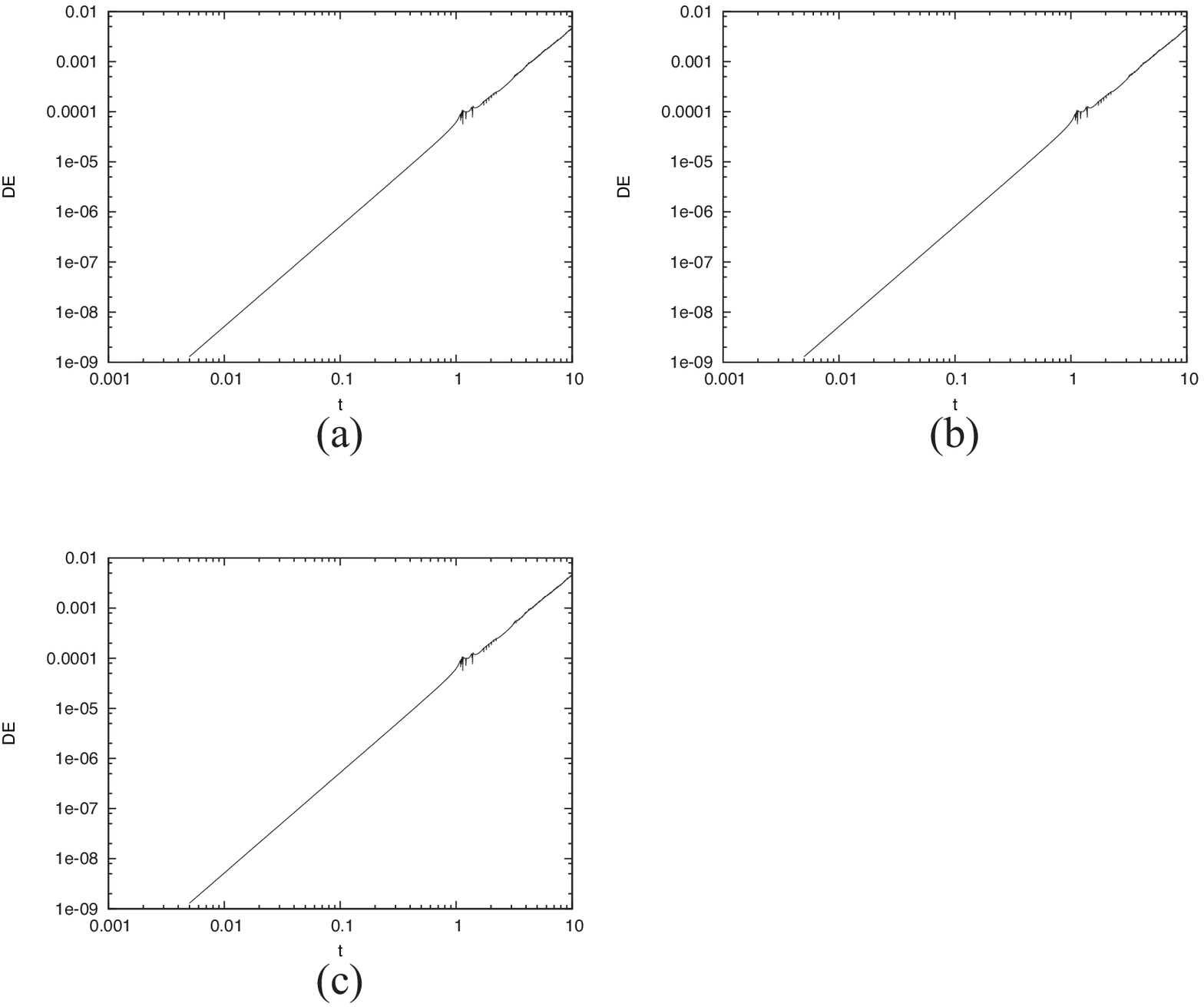}
}
\vspace{2cm}
\caption{
Error of the total energy during time evolution. 
(a) $\Delta t=2^{-17}$, (b) $\Delta t=2^{-16}$, and (c) $\Delta t=2^{-15}$.}
\label{fig:E-error}
\end{figure}

The time step is set as $\Delta t= 2^{-17}$. Here, we calculate the time evolution
until $t=10$. The error of the total energy reaches up to the 1PN order (Eq.~(\ref{eqn:energy})),
as shown in Figure \ref{fig:E-error}.
Although the total energy of the system is conserved,
because the Runge--Kutta method includes the global error, the error of the total energy
increases during the time evolution. When we change the time step
($\Delta t= 2^{-16}, 2^{-15}$), the error
of the total energy is nearly unchanged. 
At $t=10$, the error of the total energy increases to about $10^{-2}$.
The origin of the energy error can be considered as the accumulated error
because of the long time integration.

%%%% Figure %%%
%%%%%%%%%%%%%%

\begin{figure}[htb]
\centerline{
\includegraphics[height=6cm]{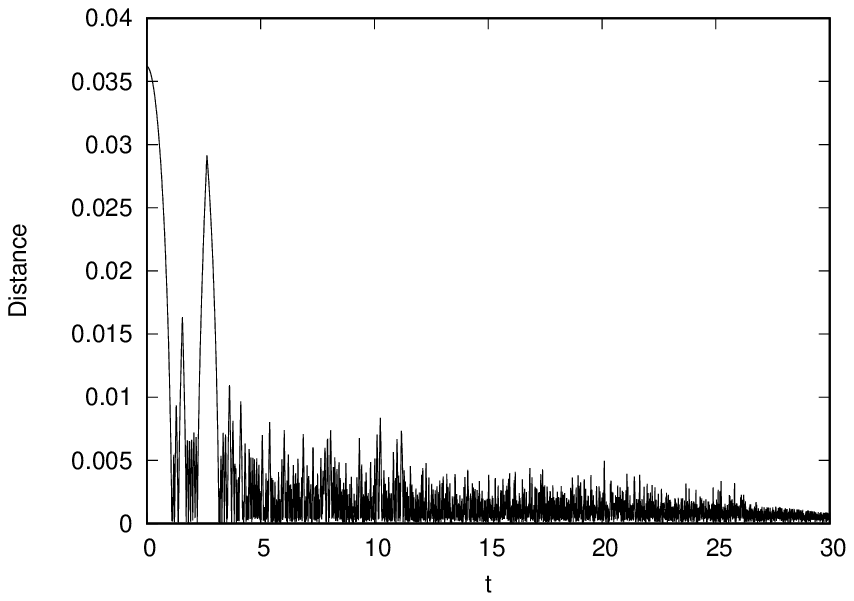}
}
\vspace{2cm}
\caption{
Distance of the nearest low-mass binary stars.
At $t \simeq 1$, the low-mass stars collide with each other. Then, the 
low-mass stars form binary systems.}
\label{fig:distance-light}
\end{figure}

\begin{figure}[htb]
\centerline{
\includegraphics[height=6cm]{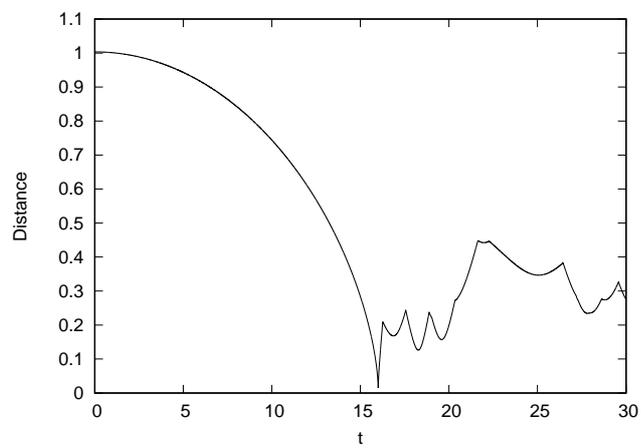}
}
\vspace{2cm}
\caption{
Distance between the massive object and nearest
low-mass star. At $t \simeq 17$, one low-mass star approaches 
the massive object. Then, the low-mass star is scattered.}
\label{fig:distance-massive}
\end{figure}

\begin{figure}[htb]
\centerline{
\includegraphics[height=6cm]{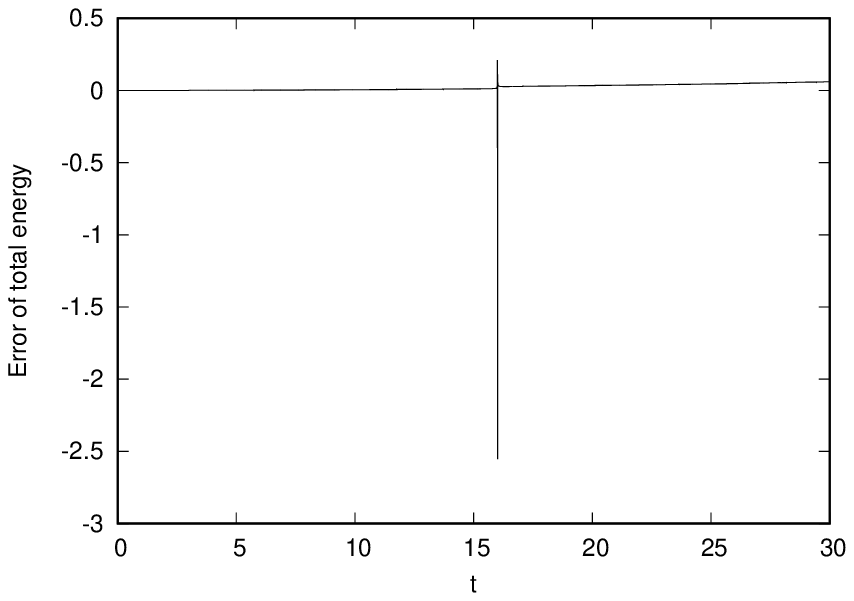}
}
\vspace{2cm}
\caption{
Error of total energy during the time evolution 
for computation time. Here, we set $\Delta t=2^{-15}$.
At $t \simeq 17$, one low-mass star approaches
the massive object. Then, a significant error appears in
the total energy. After this scattering, 
the simulation fails.}
\label{fig:E-error-t30}
\end{figure}

For analysis of the long duration, we simulate the system with $\Delta t=2^{-15}$
until $t=100$.
As is the global tendency, the stars fall to the center.
During evolution, the low-mass stars form binary systems of local clusters.
Figure~\ref{fig:distance-light} shows the distance
of the nearest low-mass binary stars. 
The low-mass stars scatter at $t \simeq 1$. According to the effect of the scattering by
low-mass stars and the formation of binaries, the error of the total energy
can oscillate. 
Figure~\ref{fig:distance-massive} shows the distance between the massive object and nearest
low-mass star. At $t \simeq 17$, one low-mass star is scattered by the massive object.
Then, the error of the total energy increases abruptly
(Fig.~\ref{fig:E-error-t30}). 
Because the simulation fails at the scattering of low-mass stars
around the massive object,
we should use a more accurate integrator such as the Hermite scheme
for the time evolution in collisional systems.

%%%% Figure %%%
%%%%%%%%%%%%%%

%%%% Figure %%%
%%%%%%%%%%%%%%

%%%% Figure %%%
%%%%%%%%%%%%%%

%%%% Figure %%%
%\begin{figure}[tb]
%\centerline{
%\includegraphics[height=7cm]{star-dist.eps}
%}
%\vspace{2cm}
%\caption{
%The relation between
%the radius and the number of light stars.
%At initial time, the stars distribute in the range of $1 < r < 10$.
%During time evolution, the stars fall to the center.}
%\label{fig:star-dist}
%\end{figure}
%%%%%%%%%%%%%%

\section{Summary}
\label{sec:summary}
In this study, an $N$-body simulation code was developed for a limited 1PN approximation.
The simulation was accelerated by a GPU, where we could conduct
realistic simulations of globular clusters with a massive object such as an intermediate-mass black hole (IMBH).

One of the critical problems in astrophysics is the conservation of total energy. Because the equation of motion
in the 1PN approximation cannot be divided into spatial terms and momentum terms, an explicit symplectic
integrator cannot be applied. As another method, the Hermite integrator was
developed~\cite{Makino1991, Makino1992, Nitadori2008}. Although
the Hermite integrator is known as a high-accuracy method, the time-derivative of the acceleration
is required. Therefore, it is difficult to apply the Hermite method to the $N$-body simulation code for the 
1PN approximation. The time derivative of the accelerations is described in Appendix~\ref{sec:a-dot}.
Because the acceleration includes three-body interactions, the time derivative of the accelerations
becomes quite complicated.

In this study, a limited 1PN approximation is considered. This code can simulate the evolution of
several astronomical objects such as globular clusters with IMBHs, accretion disks around black holes, and others.
For example, the code could simulate the accretion around Sgr A${}^*$
\cite{Gillessen2012, Ponti2015}.

As one of the scenarios of SMBH formation, the merging of many stellar-mass black holes has been
considered. For this scenario, we developed a generic 1PN approximation code with a GPU. In the generic
1PN approximation, the cost of the computation becomes $O(N^3)$. In a previous study, 
Kupi {\it et al.} proposed a direct method for time integration with the PN approximation
~\cite{Kupi2006}. Then, Brem {\it et al.} implemented the spin effects and presented a
step forward in higher-order terms~\cite{Brem2012}. In other studies using
the regularization method for the collision of stars, the higher-order PN approximation was implemented
~\cite{Aarseth2007, Harfst2008, Aarseth2012, Karl2015}. 

We implemented the 1PN approximation for the full simulation. We consider that the critical
point could be the data transfer between the GPU and main memory and the handling of shared memories 
 In previous studies using Newtonian gravity, it has been reported that a conspicuous difference in the performances 
 appears according to how the shared memories are handled~\cite{Zwart2007,Hamada2007}
Because ``Goose'' cannot be applied up to the triple loop, the command in ``Goose'' cannot
reduce the data transfer to $O(N)$. The data transfer can be reduced only to $O(N^2)$.
For the handling of shared memories and reduction of the data transfer,
we require detailed coding for a generic 1PN approximation using a CUDA C environment.

When we run this simulation, the formation process
for SMBHs from many stellar-mass black holes can be determined.
By using a GPU with a very large number of threads, the computation time is expected to reduce
considerably compared to that in general-purpose computers. 

%%%% Acknowledgments %%%%%%%%
\section*{Acknowledgments}
The author would like to thank Hideyoshi Arakida, Shuntaro Mizuno, Masahiro Morikawa, Tohru Tashiro,
and students in the astrophysics group of Ochanomizu University for their useful discussion.
The author would like to thank the roommates at the National Institute of Technology, Kochi College
for their encouragement.

\appendix
\section{Time derivation of the acceleration}
\label{sec:a-dot}
For the Hermite scheme, we require the time derivative of the acceleration for
time evolution.
The time derivative of the acceleration for massive objects can be described as follows:
%
%\begin{equation}
\begin{eqnarray}
\frac{\partial}{\partial t} \bm{a}_1 &=& 
 \frac{\partial}{\partial t} \left ( - \sum_j \frac{G m_j \bm{x}_{1j}}{r_{1j}^3} \right )
 + \frac{\partial}{\partial t} \left [ \bm{a}_1 \right ]_{BH}
 + \frac{\partial}{\partial t} \left [ \bm{a}_1 \right ]_{Cross} \,, \\
%\end{equation}
%
%\begin{eqnarray}
\frac{\partial}{\partial t} \left ( - \sum_j \frac{G m_j \bm{x}_{1j}}{r_{1j}^3} \right )
 &=& - G m_j \left ( \frac{\bm{v}_{1j}}{r_{1j}^3} - 3 \frac{(\bm{x}_{1j} \cdot \bm{v}_{1j})}{r_{1j}^5}
 \bm{x}_{1j} \right ) \,, \\
\frac{\partial}{\partial t} \left [ \bm{a}_1 \right ]_{BH}
 &=& \sum_j \frac{Gm_j}{r_{1j}^3} \left (-5 \frac{GM (\bm{x}_{1j} \cdot \bm{v}_{1j})}{r_{1j}^3}
 - 4 ( \bm{v}_j \cdot \bm{a}_j ) + \frac{3}{2} ( \bm{a}_j \cdot \bm{n}_{1j} )
 \right . \nonumber \\
 && ~~~~~~~~~~~~~~~~ \left . - \frac{3}{2} \frac{1}{r_{1j}} (\bm{v}_j \cdot \bm{x}_{1j}) (\bm{v}_{1j} \cdot \bm{x}_{1j}) 
  + \frac{3}{2} \frac{(\bm{v}_j \cdot \bm{v}_{1j})}{r_{1j}} \right ) \bm{x}_{ij} \nonumber \\
 && + \sum_j G m_j \left ( -\frac{3}{r_{1j}^5} (\bm{v}_{1j} \cdot \bm{x}_{1j} ) \bm{x}_{1j}
 + \frac{\bm{v}_{1j}}{r_{1j}^3} \right )
 \left (5 \frac{GM}{r_{1j}} - 2 v_j^2 + \frac{3}{2} (\bm{v}_j \cdot \bm{n}_{ij} ) \right ) \nonumber \\
 && - 9 \sum_j \frac{G m_j}{r_{1j}^6} (\bm{v}_{1j} \cdot \bm{x}_{1j} ) (\bm{v}_j \cdot \bm{x}_{1j} ) \bm{v}_j
 \nonumber \\
 && + 3 \sum_j \frac{G m_j}{r_{1j}^3} \left \{ (\bm{a}_j \cdot \bm{x}_{1j} ) \bm{v}_j
 + (\bm{v}_j \cdot \bm{v}_{1j}) \bm{v}_j + (\bm{v}_j \cdot \bm{x}_{1j}) \bm{a}_j \right \} \,, \\
\frac{\partial}{\partial t} \left [ \bm{a}_1 \right ]_{Cross}
 &=& 4 \sum_j G^2 m_j^2 \left ( \frac{\bm{v}_{1j}}{r_{1j}^4} - 4 \frac{(\bm{x}_{1j} \cdot \bm{v}_{1j})}{r_{1j}^6} 
 \bm{x}_{1j} \right ) \nonumber \\
 && + \sum_{j, k} G^2 m_j m_k \left ( \frac{\bm{v}_{1j}}{r_{1j}^3} 
 - 3 \frac{(\bm{x}_{1j} \cdot \bm{v}_{1j})}{r_{1j}^6} 
 \bm{x}_{1j} \right ) \left ( \frac{4}{r_{1k}} + \frac{5}{4 r_{jk}}
  -\frac{r_{1k}^2}{4 r_{jk}^3} + \frac{r_{1j}^2}{4 r_{jk}^3}
 \right ) \nonumber \\
 && + \sum_{j, k} \frac{G^2 m_j m_k \bm{x}_{1j}}{r_{1j}^3}
 \left ( -\frac{4}{r_{1k}^3} (\bm{v}_{1k} \cdot \bm{x}_{1k}) 
  - \frac{5}{4 r_{jk}^3} (\bm{x}_{jk} \cdot \bm{v}_{jk}) 
  - \frac{(\bm{x}_{1k} \cdot \bm{v}_{1k})}{2 r_{jk}^3}
  \right . \nonumber \\
 && ~~~~~~~~~~~~~~~~ \left . - \frac{3 r_{1k}^2 (\bm{x}_{jk} \cdot \bm{v}_{jk})}{4 r_{jk}^5}
  + \frac{(\bm{x}_{1j} \cdot \bm{v}_{1j})}{2r_{jk}^3}
 + \frac{3 r_{1j}^2 (\bm{x}_{jk} \cdot \bm{v}_{jk})}{4 r_{jk}^5} \right ) \nonumber \\
 && -\frac{7}{2} \sum_{j,k} \frac{G^2 m_j m_k \bm{v}_{jk}}{r_{jk}^3 r_{1j}} \nonumber \\ 
 && + \frac{7}{2} \sum_{j,k} G^2 m_j m_k \left ( \frac{3 (\bm{x}_{jk} \cdot \bm{v}_{jk})}{r_{jk}^5 r_{1j}}
 + \frac{(\bm{x}_{1j} \cdot \bm{v}_{1j})}{r_{jk}^3 r_{1j}^3} \right ) \bm{x}_{jk} \nonumber \\
 && - \sum_{j,k} \frac{G m_j m_k}{M r_{1j}^3} \left ( 4 (\bm{a}_j \cdot \bm{v}_k ) \bm{x}_{1j}
 + 4 ( \bm{v}_j \cdot \bm{a}_k ) \bm{x}_{1j} + 4 ( \bm{v}_j \cdot \bm{v}_k ) \bm{v}_{1j} \right . \nonumber \\
 && ~~~~~~~~~~~~~~~~ - 3 (\bm{a}_j \cdot \bm{x}_{1j} ) \bm{v}_k - 3 (\bm{v}_j \cdot \bm{v}_{1j}) \bm{v}_k 
   -3 (\bm{v}_j \cdot \bm{x}_{1k} ) \bm{a}_k \nonumber \\
 && ~~~~~~~~~~~~~~~~ \left . - 4 (\bm{a}_k \cdot \bm{x}_{1j} ) \bm{v}_k - 4 (\bm{v}_k \cdot \bm{v}_{1j}) \bm{v}_k 
 - 4 (\bm{v}_k \cdot \bm{x}_{1j} ) \bm{a}_k \right ) \,.
\end{eqnarray}
The time derivative of the acceleration for low-mass stars can be described as follows:
\begin{eqnarray}
\frac{\partial}{\partial t} \bm{a}_i &=& 
\frac{\partial}{\partial t} \left ( - \frac{G M \bm{x}_{j1}}{r_{j1}^3}
 - \sum_j \frac{G m_j \bm{x}_{ij}}{r_{ij}^3} \right )
 + \frac{\partial}{\partial t} \left [ \bm{a}_i \right ]_{BH}
 + \frac{\partial}{\partial t} \left [ \bm{a}_i \right ]_{Cross} \,, \\
\frac{\partial}{\partial t} \left ( - \frac{G M \bm{x}_{j1}}{r_{j1}^3}
 - \sum_j \frac{G m_j \bm{x}_{ij}}{r_{ij}^3}
 \right )
 &=& - G M \left ( \frac{\bm{v}_{j1}}{r_{j1}^3} - 3 \frac{(\bm{x}_{j1} \cdot \bm{v}_{j1})}{r_{j1}^5}
 \bm{x}_{j1} \right ) \nonumber \\
 &&  - G m_j \left ( \frac{\bm{v}_{ij}}{r_{ij}^3} - 3 \frac{(\bm{x}_{ij} \cdot \bm{v}_{ij})}{r_{ij}^5}
 \bm{x}_{ij} \right ) \,, \\
\frac{\partial}{\partial t} \left [ \bm{a}_i \right ]_{BH}
 &=& \frac{4 G^2 M^2}{r_{i1}^6} \left ( r_{i1}^2 \bm{v}_{i1} - 4 ( \bm{x}_{i1} \cdot \bm{v}_{i1} ) \bm{x}_{i1} 
 \right ) \nonumber \\
 && - \frac{GM}{r_{i1}^3} v_i^2 \bm{v}_{i1} - \frac{2GM}{r_{i1}^3} (\bm{v}_i \cdot \bm{a}_i ) \bm{x}_{i1}
 + \frac{3 GM}{r_{i1}^5} v_i^2 (\bm{x}_{i1} \cdot \bm{v}_{i1} ) \bm{x}_{i1} \,, \\
\frac{\partial}{\partial t} \left [ \bm{a}_i \right ]_{Cross}
 &=& 5 \frac{G^2 m_i M \bm{v}_{i1}}{r_{i1}^4} - 20 \frac{G^2 m_i M ( \bm{x}_{i1} \cdot \bm{v}_{i1} )}{r_{i1}^6}
 \bm{x}_{i1} \nonumber \\
 && - 3 \frac{G m_i (\bm{x}_{i1} \cdot \bm{v}_{i1})}{r_{i1}^5} \left ( 4 v_i^2 \bm{x}_{i1} - 7 (\bm{v}_i \cdot \bm{x}_{i1}
 ) \bm{v}_i \right ) \nonumber \\
 && - \frac{G m_i}{r_{i1}^3} \left [ 8 ( \bm{v}_i \cdot \bm{a}_i ) \bm{x}_{i1} + 4 v_i^2 \bm{v}_{i1} 
 - 7 \left ( (\bm{a}_i \cdot \bm{x}_{i1}) + (\bm{v}_i \cdot \bm{v}_{i1} ) \right )
 \bm{v}_i  \right . \nonumber \\
 && ~~~~~~~~ \left . - 7 ( \bm{v}_i \cdot \bm{x}_{i1} ) \bm{a}_i \right ] \nonumber \\
 && + \sum_j \frac{G^2 m_j M \bm{v}_{i1}}{r_{i1}^3}
 \left ( \frac{4}{r_{ij}} + \frac{5}{4 r_{j1}} + \frac{r_{i1}^2}{4 r_{ji}^3} - \frac{r_{ij}^2}{4 r_{j1}^3} \right ) \nonumber \\
 && -3 \sum_j \frac{G^2 m_j M (\bm{x}_{i1} \cdot \bm{v}_{i1}) \bm{v}_{i1}}{r_{i1}^5}
 \left ( \frac{4}{r_{ij}} + \frac{5}{4 r_{j1}} + \frac{r_{i1}^2}{4 r_{ji}^3} - \frac{r_{ij}^2}{4 r_{j1}^3} \right ) \nonumber \\
 && + \sum_j \frac{G^2 m_j M \bm{x}_{i1}}{r_{i1}^3}
  \left ( -\frac{4 (\bm{x}_{ij} \cdot \bm{v}_{ij} )}{r_{ij}^3} - \frac{5 (\bm{x}_{j1} \cdot \bm{v}_{j1})}{4 r_{j1}^3}
  + \frac{(\bm{x}_{i1} \cdot \bm{v}_{i1})}{2 r_{j1}^3} \right . \nonumber \\
 && ~~~~~~~~~~~~ \left . - \frac{3(\bm{x}_{j1} \cdot \bm{v}_{j1}) r_{i1}^2}{4 r_{j1}^5} 
  - \frac{(\bm{x}_{ij} \cdot \bm{v}_{ij})}{2 r_{j1}^3} + \frac{3 (\bm{x}_{j1} \cdot \bm{v}_{j1}) r_{ij}^2}{4 r_{j1}^5}
 \right ) \nonumber \\
 && + \sum_j \frac{G^2 m_j M \bm{v}_{ij}}{r_{ij}^3}
 \left ( \frac{4}{r_{ij}} + \frac{5}{4 r_{j1}} + \frac{r_{i1}^2}{4 r_{ji}^3} - \frac{r_{ij}^2}{4 r_{j1}^3} \right ) \nonumber \\
 && - 3 \sum_j \frac{G^2 m_j M (\bm{x}_{ij} \cdot \bm{v}_{ij}) \bm{v}_{ij}}{r_{ij}^5}
 \left ( \frac{4}{r_{ij}} + \frac{5}{4 r_{j1}} + \frac{r_{i1}^2}{4 r_{ji}^3} - \frac{r_{ij}^2}{4 r_{j1}^3} \right ) \nonumber \\
 && + \sum_j \frac{G^2 m_j M \bm{x}_{ij}}{r_{ij}^3}
 \left ( - \frac{4 (\bm{x}_{i1} \cdot \bm{v}_{i1} )}{r_{i1}^3} - \frac{5 (\bm{x}_{j1} \cdot \bm{v}_{j1})}{4 r_{j1}^3}
 - \frac{(\bm{x}_{i1} \cdot \bm{v}_{i1})}{2 r_{j1}^3} \right . \nonumber \\
 && ~~~~~~~~~~~~~ \left .  + \frac{3 (\bm{x}_{j1} \cdot \bm{v}_{j1} ) r_{i1}^2}{4 r_{j1}^3}
 + \frac{(\bm{x}_{ij} \cdot \bm{v}_{ij})}{2 r_{j1}^3} - \frac{3 (\bm{x}_{j1} \cdot \bm{v}_{j1}) r_{ij}^2}{4 r_{j1}^5} 
 \right ) \nonumber \\
 && - \frac{7}{2} \sum_j G^2 m_j M \left (\frac{\bm{v}_{j1}}{r_{j1}^3} 
  - \frac{3 (\bm{x}_{j1} \cdot \bm{v}_{j1} )}{r_{j1}^5} \bm{x}_{j1} \right ) \left ( \frac{1}{r_{ij}} - \frac{1}{r_{i1}} \right ) 
 \nonumber \\
 && + \frac{7}{2} \sum_j \frac{G^2 m_j M \bm{x}_{j1}}{r_{j1}^3} 
 \left ( \frac{(\bm{x}_{ij} \cdot \bm{v}_{ij})}{r_{ij}^3} - \frac{(\bm{x}_{i1} \cdot \bm{v}_{i1})}{r_{i1}^3} 
  \right ) \nonumber
\end{eqnarray}
\begin{eqnarray}
 && + \sum_j \frac{G m_j}{r_{i1}^5} (\bm{r}_{i1} \cdot \bm{v}_{i1})
 \left [ 4 (\bm{v}_i \cdot \bm{v}_j) \bm{x}_{i1} -3 (\bm{v}_j \cdot \bm{x}_{i1} ) \bm{v}_i
  -4 (\bm{v}_i \cdot \bm{x}_{i1} ) \bm{v}_j \right ] \nonumber \\
 && -\sum_j \frac{G m_j}{r_{i1}^3}
  \left [ 4 (\bm{a}_i \cdot \bm{v}_j) \bm{x}_{i1} + 4 (\bm{v}_i \cdot \bm{a}_j) \bm{x}_{i1}
 +  4 (\bm{v}_i \cdot \bm{v}_j) \bm{v}_{i1} \right . \nonumber \\
 && ~~~~~~~~ - 3 (\bm{a}_j \cdot \bm{x}_{i1}) \bm{v}_i  - 3 (\bm{v}_j \cdot \bm{v}_{i1}) \bm{v}_i 
  - 3 (\bm{v}_j \cdot \bm{x}_{i1}) \bm{a}_i \nonumber \\
 && ~~~~~~~~ \left . - 4 (\bm{a}_i \cdot \bm{x}_{i1} ) \bm{v}_j - 4 (\bm{v}_i \cdot \bm{v}_{i1} ) \bm{v}_j
  - 4 (\bm{v}_i \cdot \bm{x}_{i1} ) \bm{a}_j \right ] \nonumber \\
 && + \sum_j \frac{G m_j \bm{v}_{ij}}{r_{ij}^3}
  \left [ v_i^2 -2 | \bm{v}_{ij} |^2 + \frac{3}{2} (\bm{v}_j \cdot \bm{n}_{ij})^2 \right ] \nonumber \\
 && - \sum_j \frac{3 G m_j (\bm{x}_{ij} \cdot \bm{v}_{ij}) \bm{x}_{ij}}{r_{ij}^5}
  \left [ v_i^2 -2 | \bm{v}_{ij} |^2 + \frac{3}{2} (\bm{v}_j \cdot \bm{n}_{ij})^2 \right ] \nonumber \\
 && + \sum_j \frac{2 G m_j \bm{x}_{ij}}{r_{ij}^3}
  \left ( ( \bm{v}_i \cdot \bm{a}_i ) - 2 (\bm{v}_{ij} \cdot \bm{a}_{ij}) \right ) \nonumber \\
 && + \sum_j \frac{3 G m_j \bm{v}_{ij}}{2 r_{ij}^5} (\bm{v}_j \cdot \bm{x}_{ij} )^2
 - \frac{15 G m_j (\bm{x}_{ij} \cdot \bm{v}_{ij}) \bm{x}_{ij}}{r_{ij}^7}
  (\bm{v}_j \cdot \bm{x}_{ij} )^2 \nonumber \\
 && + \sum_j \frac{3 G m_j \bm{x}_{ij}}{r_{ij}^5}
  \left [ (\bm {a}_j \cdot \bm{x}_{ij} ) + (\bm{v}_j \cdot \bm{v}_{ij} ) \right ]
   (\bm{v}_j \cdot \bm{x}_{ij}) \nonumber \\
 && - \sum_j \frac{3G m_j (\bm{x}_{ij} \cdot \bm{v}_{ij})}{r_{ij}^5}
 \left [ \bm{x}_{ij} \cdot (4 \bm{v}_i - 3 \bm{v}_j ) \right ] \bm{v}_{ij} \nonumber \\
 && + \sum_j \frac{G m_j}{r_{ij}^3} \left [ \left \{ \bm{v}_{ij} \cdot (4 \bm{v}_i - 3 \bm{v}_j )
 \right \} + \left \{ \bm{x}_{ij} \cdot (4 \bm{a}_i - 3 \bm{a}_j ) \right \} \right ] \bm{v}_{ij}
  \nonumber \\
 && + \sum_j \frac{G m_j}{r_{ij}^3} 
 \left [ \bm{x}_{ij} \cdot (4 \bm{v}_i - 3 \bm{v}_j ) \right ] \bm{a}_{ij} \,.
\end{eqnarray}
In these equations, the acceleration $\bm{a}_1, \bm{a}_i$ has already been described in
Eqs.~(\ref{eqn:eq-object1}) and (\ref{eqn:eq-light_object}).

%%%% Bibliography  %%%%%%%%%%

\end{document}